\documentclass[journal]{IEEEtran}
\ifCLASSINFOpdf

\else

\fi

\usepackage{cite}
\usepackage{amsmath,amssymb,amsfonts}
\usepackage{graphicx}
\usepackage{textcomp}
\usepackage{xcolor}
\usepackage{caption}
\usepackage{mathtools}
\usepackage{amsfonts}
\usepackage{verbatim}
\usepackage{stfloats}
\usepackage{subfig}
\usepackage{multirow}
\usepackage{bbm}

\usepackage{algorithm}
\usepackage{algorithmic}

\usepackage{amsmath,amssymb,amsthm}

\newtheorem{definition}{Definition}
\newtheorem{lemma}{Lemma}
\newtheorem{proposition}{Proposition}

\def\BibTeX{{\rm B\kern-.05em{\sc i\kern-.025em b}\kern-.08em
    T\kern-.1667em\lower.7ex\hbox{E}\kern-.125emX}}

\begin{document}

{\title{Value-Driven Agentic ISAC: From User Intent to Autonomous Service Provisioning}}
\author{Biwei Li,~\IEEEmembership{Member,~IEEE}   \IEEEcompsocitemizethanks{\IEEEcompsocthanksitem B. Li is  with the School of Computer Science and Engineering, Southeast University, Nanjing, 210096, China (e-mail: biwei.li@seu.edu.cn).}}
%
%


\maketitle

\begin{abstract}
The increasing autonomy of intelligent applications poses a challenge for integrated sensing and communication (ISAC) networks. Communication and sensing requirements change with task progress and environmental conditions, making service goals difficult to specify in advance. Conventional resource allocation methods typically assume predefined requirements, limiting their ability to adapt to evolving task needs. We propose an agent-in-the-loop ISAC framework for autonomous service-goal determination and adaptive resource allocation. At each decision epoch, an agent translates the user's high-level intent into a structured service goal using its current task state and environmental knowledge. The goal specifies communication demand and latency limits, sensing targets and accuracy requirements, and their task-dependent importance. It provides an explicit interface between agent-level decisions and network resource management. The network evaluates how well communication and sensing services fulfill the current goals and jointly allocates subchannels and transmit power to maximize the system value of service. Communication outcomes and sensing observations inform the agent's service goals for the next epoch. We develop a value-driven resource allocation method based on a Stackelberg mechanism that accounts for the benefits of shared sensing. Numerical experiments assess system value of service, communication and sensing resource use, and adaptation to changing service goals.
\end{abstract}

\begin{IEEEkeywords}
Agentic communication networks, integrated sensing and communication (ISAC),
user intent, value-driven resource allocation, autonomous service provisioning.
\end{IEEEkeywords}

\section{Introduction}

Future wireless networks are expected to support increasingly
autonomous and intelligent services, whose communication and
sensing requirements evolve with task objectives and environmental
conditions \cite{11419150}. Integrated sensing and communication (ISAC) provides
essential capabilities for such services by enabling information
exchange and environmental sensing through shared wireless
resources \cite{11250835}. However, autonomous task execution requires ISAC
networks to adapt their operations to the decisions and information
needs of intelligent agents. Unlike conventional services with
predefined quality-of-service (QoS) requirements, agent-assisted
services may begin with only a high-level user intent, without
specifying the intermediate objectives or network services needed
for its execution. These objectives must be determined dynamically
based on task states, environmental knowledge, and newly acquired
observations. Consequently, network resource management must
accommodate communication and sensing requirements that evolve
with environmental information.

To support such evolving requirements, ISAC networks need
continuous interaction between intelligent agents and resource
management \cite{11358925}. Sensing observations
update the agent's environmental knowledge, while communication
supports information exchange for task execution. Based on its
current task state and knowledge, the agent determines the
required information and corresponding communication and sensing
goals. The network allocates radio resources accordingly, and
the resulting observations and service outcomes inform subsequent
agent decisions. The agent uses this feedback to update the service goals that guide the next round of resource allocation.

This interaction is particularly important 
in dynamic environments. Consider an autonomous mobile service
whose user intent is to reach a destination safely and efficiently.
During normal operation, coarse environmental information may
suffice for navigation. However, a moving obstacle may introduce
a new sensing objective requiring more accurate information
about its position and motion. As the obstacle is identified
and its uncertainty decreases, further sensing may become
less necessary, allowing resources to be redirected toward
other task-relevant operations. Throughout this process,
the user intent remains unchanged, while intermediate goals
and resource requirements evolve with the agent's knowledge
and task state. Thus, the utility of communication and sensing
resources depends on both conventional performance and their
contribution to the current task objective.

Supporting such adaptive services introduces three fundamental
challenges for resource management in agentic ISAC networks.

\textit{1) Dynamic service-goal determination from
	high-level user intents:}
A high-level intent does not specify the communication and
sensing requirements throughout task execution.
The agent need to determine service goals from its
task state and environmental knowledge, and update them
as new observations become available. Existing ISAC resource
allocation methods typically assume predefined requirements
\cite{10979960}, without explicitly accounting for
how acquired information changes subsequent service goals.
A policy is therefore needed to connect
user intents with dynamically generated requirements and
establish an interface between agent decisions and network
resource management.

\textit{2) Task-dependent evaluation metric of
	communication and sensing services:}
The contribution of wireless resources to task execution
depends on the current objectives.
Sensing value varies with target relevance, state uncertainty,
and required accuracy, while communication value depends
on the requested data rate and amount \cite{11184506}.
Moreover, multiple agents may require information about
common targets, creating opportunities for shared sensing.
Conventional metrics such as communication rate and sensing
accuracy alone cannot fully capture these dependencies.
A value of service model is needed to quantify the
contribution of heterogeneous services to intent execution.

\textit{3) Adaptive resource allocation under
	dynamic service requests:}
Resource allocation affects both immediate service performance
and the information available for subsequent agent decisions.
As new observations update environmental knowledge, service
goals and resource requirements may change accordingly.
Instantaneous performance optimization may overlook these
dependencies, leading to redundant information acquisition
or insufficient support for emerging requirements. Meanwhile,
multiple agents compete for limited radio resources. An adaptive allocation mechanism is therefore needed to coordinate communication and sensing
according to task-dependent values while accommodating goal
updates and shared resource constraints.

\subsection{Related Work}

\subsubsection{Agentic AI and Agentic Communication Networks}

Agentic artificial intelligence (AI) has emerged as a framework for autonomous context
interpretation, reasoning, goal-oriented decision-making,
and feedback-based adaptation through continuous
perception--reasoning--action interactions
\cite{11339915,11519591,jiang2026agentic}. In wireless
networks, recent studies have explored intent interpretation,
autonomous management, agent collaboration, and communication
support for distributed agents \cite{li2026agentic}. The authors in \cite{11535442} proposed an agentic
framework for UAV-assisted ambient Internet of Things (IoT) that uses task feedback
to guide distributed learning for cooperative backscatter sensing. Intent-driven frameworks further translate natural-language requests into executable network control strategies by leveraging intelligent reasoning modules to interpret user objectives and real-time feedback to adapt to dynamic conditions \cite{11339899}.

Agentic AI has also attracted attention in ISAC systems,
where large language models (LLMs), generative models, deep reinforcement learning,
and mixtures of experts have been employed to improve
environmental sensing, reward design, reasoning, and network
control \cite{11673126,acm124}. These approaches demonstrate
the potential of agent-based adaptation, but commonly retain
predefined communication and sensing objectives, such as
maximizing data rate or minimizing sensing error. 
\subsubsection{Dynamic and Task-Oriented ISAC Resource Allocation}

Joint ISAC resource allocation has been widely investigated
to balance communication and sensing requirements under
limited spectrum, power, and time resources
\cite{11271832,11397292}. Existing studies optimize power,
beamforming, bandwidth, and scheduling for communication
rate, sensing accuracy, energy efficiency, and service-oriented
performance. For a given set of sensing targets, the MEC-assisted ISAC design
in~\cite{11175184} jointly optimizes sensing scheduling, beamforming
for sensing and data offloading, and computation resource allocation
under task latency and model training error constraints. Dynamic designs further account for mobility,
moving targets, time-varying channels, and environmental
uncertainty \cite{11520265,11614187}. However, resource
adaptation generally operates under predefined sensing
targets, or performance requirements.

Mason et al.~\cite{10490134} used multi-agent reinforcement
learning to jointly optimize communication and control, prioritizing
transmissions based on the relevance of the information to the control task.
Task-oriented communication and information-aware sensing
prioritize information according to its usefulness for
downstream inference and decision tasks
\cite{11395628,11610640}. These approaches recognize
heterogeneous information value and reduce unnecessary
acquisition through task relevance. Nevertheless, the
downstream task or information objective is generally
specified in advance. In agent-driven services, new
observations may change the intermediate goal, the relevant
physical states, and their required sensing accuracy.

\subsection{Contributions}
To address these challenges, this paper proposes a value-driven agentic ISAC framework that integrates dynamic service-goal generation with adaptive network resource allocation. In this framework, each agent translates the user's intent into structured communication and sensing requirements based on its task state and environmental knowledge, while the network allocates radio resources according to the value of fulfilling these requirements. The resulting service outcomes are fed back to the agents to support subsequent goal updates, establishing a closed-loop interaction between agent-level decision-making and network-level resource management. The main contributions are summarized as follows.

1) We introduce an agent-in-the-loop service model that
separates persistent user intents from state-dependent
intermediate goals. Each agent translates its intent,
task state, and environmental posterior information into structured
communication and sensing requirements, including
communication demand, sensing targets, accuracy requirements,
and target relevance. ISAC observations subsequently
update the agent's posterior information and task state, enabling
continuous goal regeneration through a closed
intent--goal--resource--observation loop.

2) \textcolor{black}{We propose an intent-aware value model for current service
goals. The model quantifies communication and sensing
fulfillment while accounting for target relevance and available
sensing information.} For targets
requested by multiple agents, individual sensing values
are aggregated to account for shared information acquisition.
\textcolor{black}{Based on this model, we maximize the system value of service
(VoS) at each decision epoch by optimizing communication and sensing
subchannel assignments and transmit powers for the current goals.}

3) \textcolor{black}{We develop a value-driven online allocation algorithm based on
a Stackelberg mechanism. We derive supporting prices that account
for the marginal benefits of shared sensing to other agents. The
algorithm searches a discrete neighborhood of subchannel assignments
and optimizes transmit power for each candidate.} New ISAC observations
are then fed back to the agents for subsequent goal updates.

The remainder of this paper is organized as follows.
Section II presents the system model and intent--goal
evolution. \textcolor{black}{Section~\ref{sec:methodology} develops the Stackelberg
mechanism and the online allocation algorithm. Section~ IV
reports the numerical simulation results, and Section~\ref{sec:conclusion} concludes
the paper.}
\begin{figure*}[!t]
	\centering
	\includegraphics[width=0.9\textwidth]{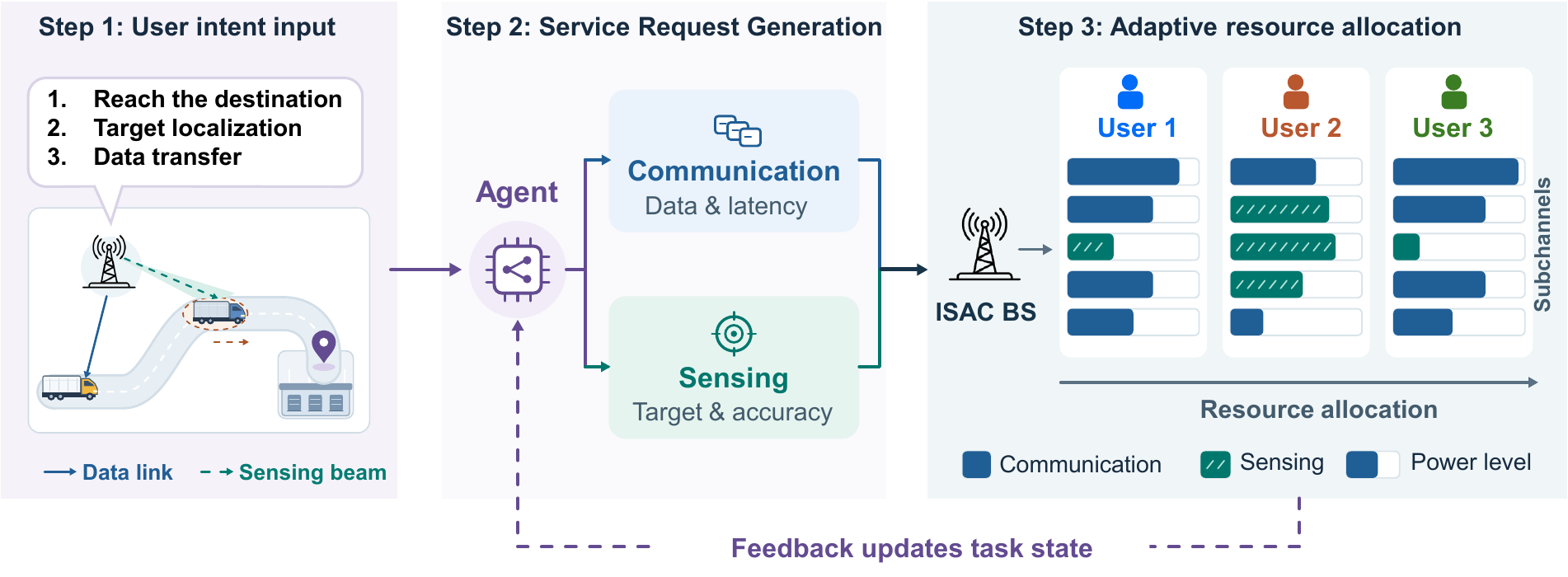}
	\caption{The architecture of the multi-user agentic ISAC system. The agent interprets the user's intent into a structured service request that specifies the data rate, latency limit, sensing targets, and required sensing accuracy. The base station uses these request to allocate subchannels and transmit power across users. Communication outcomes and sensing observations inform the agents' service request for the next decision epoch.}
	\label{fig0}
\end{figure*}
\vspace{-5mm}
\section{System Model}
\label{sec:system_model}
\label{subsec:goal_generation}
\subsection{System Architecture}
In the considered ISAC system, there is one base station (BS) serving $K$ users with various intents. Each user \(k\) conveys a high-level service objective to its associated agent through an intent \(\mathcal{I}_k\). At each decision epoch, agent \(k\) interprets the intent of the user according to the current task state \(\mathbf{x}_k^t\) and the predictive position distributions collected in \(\mathbf{b}_k^t\), and converts it into a structured service goal \(\mathbf{g}_k^t=\mathcal{G}_k(\mathcal{I}_k,\mathbf{x}_k^t,\mathbf{b}_k^t)\). For each tracked target, \(\mathbf{b}_k^t\) contains its ID, class label, and the mean and covariance of its predictive position distribution in \eqref{eq:sensing_prediction_prior}. These distributions are available at the start of epoch \(t\) and are obtained by propagating the previous epoch's filtering posteriors. The service goal for a fixed intent can therefore change as the task state or predictive distributions change.  In this paper, we introduce the concepts of \textit{intent}, \textit{agent}, and \textit{service request} as


\begin{itemize}
	
\item \textit{Agent.} The decision-making entity that generates a service request for user \(k\) from the user's intent, current task state, and predictive target-position distributions.

  \item \textit{Intent.} An intent \(\mathcal{I}_k\) is a list of clauses,
each pairing a trigger with an action. A trigger activates the clause
every epoch, at a specified interval, or upon a specified event type.
A \texttt{SENSE} action specifies a region, an object class, and a profile
defining the target limit, relevance scoring rule, localization tolerance,
and priority. A \texttt{SEND} action specifies a data product and a profile
defining the delivery time budget from packet release. The agent encodes
the data product into packets, each with an initial size equal to its
encoded bit length.


	\item \textit{Service Request.} The structured specification \(\mathbf{g}_k^t\) of communication and sensing requirements and their relative priorities, submitted to the BS at decision epoch \(t\).
\end{itemize}

As illustrated in Fig. 1, consider a vehicle delivery task in which the user intends to reach a destination, localize a nearby vehicle, and transfer data. Rather than directly specifying communication and sensing parameters, the user provides these high-level intents to its agent. At each decision epoch, the agent evaluates the current task progress and the predictive position distributions to determine the corresponding service goal, including the data volume and delivery deadline for communication, as well as the target and required localization accuracy for sensing. These requirements may evolve as the task progresses. For instance, approaching the destination may call for more accurate localization, while an approaching data delivery deadline may increase the urgency of communication. The BS then allocates subchannels and transmit power according to the resulting requirements. The communication and sensing outcomes update the task state and predictive position distributions, respectively, allowing the agent to adapt its service goal at the next decision epoch.

\vspace{-4mm}
\subsection{Structured service request}
We introduce structured service request to capture four aspects that determine the utility of network resources to the current task: communication demand and urgency, sensing objects and required accuracy, the relative importance between communication and sensing, and the task relevance of different sensing targets. Thus, the structured service request of agent $k$ at time slot $t$ is represented by
\begin{align}
	\mathbf{g}_k^t
	=
	\left\{
	D_k^t,\,
	L_k^t,\,
	\alpha_k^t,\,
	\mathcal{M}_k^t,\,
	\boldsymbol{\epsilon}_k^t,\,
	\boldsymbol{\phi}_k^t
	\right\}.
	\label{eq:goal_definition}
\end{align}
For the communication, \(D_k^t\) denotes the amount of data to be received by the user, while \(L_k^t\) denotes the corresponding delay request. Therefore, the required communication rate is
\begin{align}
	R_{k,\mathrm{req}}^t
	=
	\frac{D_k^t}{L_k^t}.
	\label{eq:required_rate_system}
\end{align}
The coefficient \(\alpha_k^t\in[0,1]\) describes the relative importance of communication and sensing for the current goal, where a larger \(\alpha_k^t\) places greater emphasis on communication and \(1-\alpha_k^t\) on sensing. The coefficient may vary across epochs under the same intent as the set of requested services changes.

For the sensing component of the goal, the agent first determines which physical targets are relevant to the current task. We denote this target set by $\mathcal{M}_k^t \subseteq \mathcal{M}$. For each selected target $m \in \mathcal{M}_k^t$, the agent specifies a required sensing accuracy $\epsilon_{k,m}^t$, which represents the maximum tolerable localization error for that target. Different sensing targets may contribute differently to the current task. We therefore associate each selected target with a relevance weight $\phi_{k,m}^t \geq 0$, where a larger $\phi_{k,m}^t$ indicates that accurate knowledge of target $m$ is more important for accomplishing the current goal. The relevance weights are normalized such that
$\sum_{m \in \mathcal{M}_k^t} \phi_{k,m}^t = 1.$. When the current request does not require sensing, we set $\mathcal{M}_k^t=\emptyset$ and $\alpha_k^t=1$, in which case no target-relevance weights are defined. Therefore, the normalization applies only when $\mathcal{M}_k^t\neq\emptyset$. Conversely, when sensing is required, $\alpha_k^t<1$ and the relevance weights characterize the relative importance of the selected sensing targets.
\subsection{From intent to service request}
We implement $\mathcal G_k$ as a deterministic mapping from an intent
expressed as a list of clauses $\mathcal I_k=((\gamma_r,a_r))_{r=1}^{N_k}$.
Each clause pairs one trigger $\gamma_r$ with one action $a_r$, either
\texttt{SENSE}(region, class, profile) or \texttt{SEND}(product, profile).
The application defines the selectable regions, object classes, event
types, data products, and profiles. The task state $\mathbf x_k^t$
provides newly available events, data products, and the pending packet
queue. The agent uses the predictive position distributions in
$\mathbf b_k^t$ to generate sensing requests. 


\textit{1) Activate clauses.}
\texttt{EVERY\_EPOCH} activates at every decision epoch.
\texttt{PERIODIC}$(P_r)$ activates every $P_r$ epochs from its configured
start epoch. \texttt{ON\_EVENT}$(e)$ activates once for each newly
available event of type $e$. Index each resulting execution by $h$.
An instance $h$ inherits the action and profile of its clause, and
multiple events can produce multiple instances of the same clause.
Let $\mathcal A_k^{s,t}$ denote the active sensing instances.

\textit{2) Generate sensing requirements.}
Each active sensing instance $h$ filters the tracks in $\mathbf b_k^t$
by region, using the predictive means, and by object class. Its profile specifies a target limit $B_h$,
a rule for computing positive relevance scores $s_{h,m}^t$, a
localization tolerance $\epsilon_h>0$, and a priority $p_h>0$.
For example, a proximity-based profile uses
$s_{h,m}^t=(1+d_{h,m}^t/d_0)^{-1}$, where $d_{h,m}^t$ is the predicted
distance to the monitoring reference point and $d_0>0$ is a distance
scale. Let $\mathcal S_h^t$ contain up to $B_h$ eligible tracks with the
highest scores, with ties broken by target ID. The agent merges
duplicate targets and retains the strictest tolerance

\begin{align}
	\mathcal M_k^t &= \bigcup_{h\in\mathcal A_k^{s,t}}\mathcal S_h^t,\\
	\epsilon_{k,m}^t &= \min_{h:\,m\in\mathcal S_h^t}\epsilon_h,\\
	w_{k,m}^t &= \sum_{h:\,m\in\mathcal S_h^t}
	p_h\frac{s_{h,m}^t}{\sum_{j\in\mathcal S_h^t}s_{h,j}^t},\\
	\phi_{k,m}^t &= \frac{w_{k,m}^t}{\sum_{j\in\mathcal M_k^t}w_{k,j}^t},
	\qquad m\in\mathcal M_k^t.
\end{align}

The minimum and sum range over active sensing instances that select
target $m$. Each nonempty selection contributes total weight $p_h$.

\textit{3) Generate communication requirements.}
Each active communication instance encodes the specified data product
if available and enqueues the resulting packets for delivery to user $k$.
The initial remaining length of packet $i$ is its encoded bit length.
A packet released at time $u_i$ by instance $h$ has absolute deadline
$d_i=u_i+\Delta_h$, where $\Delta_h>0$ is the profile's delivery budget.
At request time $\tau^t$, let $\mathcal P_k^t$ contain all queued packets
with $u_i\leq\tau^t<d_i$ and remaining length $\ell_i^t>0$, including
unfinished packets from earlier epochs. If $\mathcal P_k^t$ is nonempty,
the agent computes
\begin{equation}
	D_k^t=\sum_{i\in\mathcal P_k^t}\ell_i^t,\qquad
	L_k^t=\min_{i\in\mathcal P_k^t}(d_i-\tau^t).
\end{equation}
This aggregation assigns the earliest deadline to the entire batch.

\textit{4) Assemble the request and update the queue.}
For joint requests, $\alpha_k^t=\bar\alpha_k\in(0,1)$ is the
application preference. For communication-only requests,
$\alpha_k^t=1$. For sensing-only requests, $\alpha_k^t=0$, $D_k^t=0$,
and $L_k^t=\tau^{t+1}-\tau^t$. Thus, $\alpha_k^t$ can vary across
epochs under the same intent as the set of requested services changes.
The agent submits \eqref{eq:goal_definition} if either service is
requested.

Within the allocated communication service, the BS serves unexpired
packets in earliest-deadline order, with ties broken by enqueue order.
Let $c_i^t$ be the feedback on bits of packet $i$ delivered during
epoch $t$ by its deadline. Before the next request, the agent sets
$\ell_i^{t+1}=\ell_i^t-c_i^t$ and retains the packet with its original
deadline only if $\ell_i^{t+1}>0$ and $d_i>\tau^{t+1}$.

\subsection{Communication and Sensing Model}
The total bandwidth $B$ is divided into $N$ orthogonal subchannels, indexed by
$\mathcal{N}=\{1,2,\ldots,N\}$, each with bandwidth $\Delta f$, such that
$B=N\Delta f$. At decision epoch $t$, we define the binary variables
$\chi_{k,n}^{c,t}\in\{0,1\}$ and $\chi_{m,n}^{s,t}\in\{0,1\}$ for communication and sensing subchannel allocation, respectively. Specifically, $\chi_{k,n}^{c,t}=1$ indicates that subchannel $n$ is assigned to agent $k$ for communication, while $\chi_{m,n}^{s,t}=1$ indicates that the same subchannel is assigned to sensing target $m$. Each subchannel can support at most one communication transmission or one sensing operation within a resource-allocation interval, i.e.,
$\sum_{k=1}^{K}\chi_{k,n}^{c,t}
	+
	\sum_{m=1}^{M}\chi_{m,n}^{s,t}
	\leq 1,
	\quad
	\forall n\in\mathcal{N},\; t\in\mathcal{T}.$ Let $p_{k,n}^{c,t}\geq0$ and $p_{m,n}^{s,t}\geq0$ denote the transmit powers allocated to the communication transmission of agent $k$ and the sensing operation of target $m$ on subchannel $n$, respectively. The total transmit power at each decision epoch is constrained by
$	\sum_{n=1}^{N}
	\left(
	\sum_{k=1}^{K}
	\chi_{k,n}^{c,t}p_{k,n}^{c,t}
	+
	\sum_{m=1}^{M}
	\chi_{m,n}^{s,t}p_{m,n}^{s,t}
	\right)
	\leq
	P_{\max},
	\quad
	\forall t\in\mathcal{T}.$

\subsubsection{Communication Model}
On subchannel $n$, let $\mathbf h_{k,n}^t\in\mathbb C^{N_{\rm BS}\times1}$
and $\mathbf w_{k,n}^{c,t}$ be the downlink channel and unit-norm
transmit beamformer. The received symbol is
$y_{k,n}^{c,t}=\sqrt{p_{k,n}^{c,t}}(\mathbf h_{k,n}^t)^H
\mathbf w_{k,n}^{c,t}s_{k,n}^{c,t}+n_{k,n}^{c,t}$, where
$s_{k,n}^{c,t}$ has unit power and
$n_{k,n}^{c,t}\sim\mathcal{CN}(0,\sigma_c^2)$. With
$\gamma_{k,n}^{c,t}=p_{k,n}^{c,t}|(\mathbf h_{k,n}^t)^H
\mathbf w_{k,n}^{c,t}|^2/\sigma_c^2$, the achievable rate and available
transmission volume over an epoch of duration $\Delta T$ are
\begin{align}
	R_k^t &= \sum_{n=1}^{N}\chi_{k,n}^{c,t}\Delta f
	\log_2(1+\gamma_{k,n}^{c,t}),\label{eq:communication_rate}\\
	C_k^t &= R_k^t\Delta T.\label{eq:delivered_data}
\end{align}

\subsubsection{Sensing Model}
Let $\mathbf q_m^t=[x_m^t,y_m^t]^T$ denote the position of target $m$,
assumed approximately constant over the sensing interval. It evolves as
\begin{equation}
	\mathbf q_m^{t+1}=\mathcal F_m(\mathbf q_m^t)+\mathbf v_m^t,
	\label{eq:target_dynamics}
\end{equation}
where $\mathcal F_m$ is the position transition and $\mathbf v_m^t$ is
zero-mean process noise. Before sensing, the BS propagates the previous
filtering posterior to obtain the predictive position distribution
\begin{equation}
	\mathbf q_m^t\sim\mathcal N\!\left(
	\widehat{\mathbf q}_m^{t|t-1},\mathbf\Sigma_m^{t|t-1}\right),
	\qquad \mathbf\Sigma_m^{t|t-1}\succ\mathbf 0.
	\label{eq:sensing_prediction_prior}
\end{equation}
This distribution is the prior for the epoch-$t$ sensing update, which
produces the filtered estimate $\widehat{\mathbf q}_m^t$. We assume
initialized tracks with finite predictive covariance and use a Gaussian
approximation for nonlinear motion.

For position $\mathbf q$, let $\theta_m(\mathbf q)$ be the direction from
the BS and $\tau_m(\mathbf q)=2\|\mathbf q-\mathbf q_{\rm BS}\|/c_0$ the
round-trip delay, where $\mathbf q_{\rm BS}$ is the BS position and $c_0$
is the propagation speed. The OFDM delay phase follows~\cite{11271832}.
On tone $\ell\in\mathcal L_n$ of a sensing subchannel assigned to target
$m$, the BS observes
\begin{align}
	\mathbf y_{m,n,\ell}^{s,t}
	&=\sqrt{p_{m,n}^{s,t}}\,\alpha_{m,n}^t
	e^{-j2\pi f_{n,\ell}\tau_m(\mathbf q_m^t)}\notag\\
	&\quad\times\mathbf a_r(\theta)\mathbf a_t^H(\theta)
	\mathbf w_{m,n}^{s,t}s_{m,n,\ell}^{s,t}+\mathbf n_{m,n,\ell}^{s,t},
	\label{eq:sensing_signal}
\end{align}
where $\theta=\theta_m(\mathbf q_m^t)$, $f_{n,\ell}$ is the tone frequency,
and $\mathbf a_r$ and $\mathbf a_t$ are the receive and transmit array
responses. The effective round-trip coefficient $\alpha_{m,n}^t$,
beamformer $\mathbf w_{m,n}^{s,t}$, and pilots are treated as known during
allocation. The pilots satisfy $\sum_{\ell\in\mathcal L_n}|s_{m,n,\ell}^{s,t}|^2=1$,
so $p_{m,n}^{s,t}$ is the subchannel power. The noise is circular complex
Gaussian with covariance $\sigma_s^2\mathbf I$ independent of position and
power, and is independent across tones and assigned subchannels.

Stacking the tones gives $\mathbf y_{m,n}^{s,t}=\sqrt{p_{m,n}^{s,t}}\,
\mathbf u_{m,n}^t(\mathbf q_m^t)+\mathbf n_{m,n}^{s,t}$, with unit-power
mean $\mathbf u_{m,n}^t(\mathbf q)$ from \eqref{eq:sensing_signal}.
The BS linearizes this mean at $\widehat{\mathbf q}_m^{t|t-1}$. Let
$\mathbf D_{m,n}^t$ be $\partial\mathbf u_{m,n}^t/\partial\mathbf q^T$
evaluated at that position. The local Gaussian likelihood yields
measurement Fisher information $p\mathbf H_{m,n}^t$ at power $p$, where
$\mathbf H_{m,n}^t=2\operatorname{Re}\{(\mathbf D_{m,n}^t)^H
\mathbf D_{m,n}^t\}/\sigma_s^2\succeq\mathbf 0$ is fixed during power
optimization and may be rank deficient. Adding independent measurement
information to the predictive prior gives
\begin{align}
	\mathbf J_m^t &= (\mathbf\Sigma_m^{t|t-1})^{-1}
	+\sum_{n=1}^{N}\chi_{m,n}^{s,t}p_{m,n}^{s,t}\mathbf H_{m,n}^t,
	\label{eq:sensing_fim}\\
	\mathrm{CRLB}_m^t &=
	\sqrt{\operatorname{tr}\bigl[(\mathbf J_m^t)^{-1}\bigr]}.
	\label{eq:sensing_crlb}
\end{align}
The prior makes $\mathbf J_m^t$ positive definite for every feasible
allocation. With zero new sensing power, $\mathrm{CRLB}_m^t$ reduces
to $\sqrt{\operatorname{tr}\mathbf\Sigma_m^{t|t-1}}$. It is a root-mean-square
position error bound in the local linear Gaussian model and a quality
proxy for the nonlinear model. The BS maintains the updated estimate
$\widehat{\mathbf q}_m^t$ and its uncertainty per physical target, so
agents requesting the same target share the sensing result while
evaluating it against their individual accuracy requirements and
relevance weights.

\subsection{Agentic Feedback Model}
\label{subsec:agentic_feedback}
Let $[\tau^t,\tau^{t+1})$ denote epoch $t$, with duration $\Delta T$.
The agent generates $\mathbf g_k^t$ at $\tau^t$, and the BS executes
the resulting allocation during the epoch. The agent incorporates the
communication and sensing outcomes before generating the next request.

\textit{1) Communication feedback and task progress.}
The rate $R_k^t$ in \eqref{eq:communication_rate} is modeled as constant
within the epoch. The agent enqueues newly generated packets at the
epoch boundary, and the BS serves only the eligible packet set
$\mathcal P_k^t$ fixed at $\tau^t$. It uses earliest-deadline-first
service, with ties broken by enqueue order. For a nonempty queue and
$R_k^t>0$, index the packets in this service order and let $s_i^t$
denote the earliest time at which service of packet $i$ can begin.
Under ideal bit-level service,
\begin{align}
	s_1^t &= \tau^t, \qquad b_i^t=\min\{d_i,\tau^{t+1}\},\notag\\
	c_i^t &= \min\!\left\{\ell_i^t,
	R_k^t[b_i^t-s_i^t]_+\right\},\notag\\
	s_{i+1}^t &= s_i^t+c_i^t/R_k^t,
	\label{eq:feedback_packet_service}
\end{align}
where $[z]_+=\max\{z,0\}$ and $c_i^t$ is the amount of packet $i$
delivered by its deadline during the epoch. If $R_k^t=0$, all
$c_i^t$ are zero. The communication feedback and aggregate delivery are
\begin{align}
	\mathcal O_k^{c,t} &= \{(i,c_i^t):i\in\mathcal P_k^t\},\notag\\
	A_k^t &= \sum_{i\in\mathcal P_k^t}c_i^t
	\leq \min\{D_k^t,C_k^t\},
	\label{eq:feedback_delivered_bits}
\end{align}
where $C_k^t$ in \eqref{eq:delivered_data} is the available transmission
volume for the epoch. The application receives the decoded payload of
each packet completed by its deadline. The agent carries forward
unfinished packets as
\begin{equation}
	\begin{split}
		\mathcal Q_k^{t+1,-}=\{(i,u_i,d_i,\ell_i^t-c_i^t):\;&
		i\in\mathcal P_k^t,\\[-2pt]
		&\ell_i^t>c_i^t,\ d_i>\tau^{t+1}\}.
	\end{split}
	\label{eq:feedback_queue_update}
\end{equation}
The superscript $-$ denotes the queue before new arrivals at
$\tau^{t+1}$. Surviving packets retain their enqueue order and original
deadlines. Their remaining lengths and deadlines enter the next request.

\textit{2) Sensing feedback and predictive distribution update.}
The BS updates the position distribution of each physical target.
Under the local linear Gaussian model used in \eqref{eq:sensing_fim},
the filtering posterior is
$\pi_m^{t|t}=\mathcal N(\widehat{\mathbf q}_m^t,\mathbf P_m^{t|t})$,
with the following information-form update:
\begin{align}
	\mathbf e_{m,n}^t
	&=\mathbf y_{m,n}^{s,t}
	-\sqrt{p_{m,n}^{s,t}}\,
	\mathbf u_{m,n}^t(\widehat{\mathbf q}_m^{t|t-1}),\notag\\
	\mathbf z_m^t
	&=\sum_{n:\,\chi_{m,n}^{s,t}=1}
	\frac{2\sqrt{p_{m,n}^{s,t}}}{\sigma_s^2}
	\operatorname{Re}\!\left\{(\mathbf D_{m,n}^t)^H
	\mathbf e_{m,n}^t\right\},\notag\\
	\mathbf P_m^{t|t}&=(\mathbf J_m^t)^{-1},\notag\\
	\widehat{\mathbf q}_m^t
	&=\widehat{\mathbf q}_m^{t|t-1}+\mathbf P_m^{t|t}\mathbf z_m^t.
	\label{eq:feedback_posterior}
\end{align}
Here $\mathbf D_{m,n}^t$ is evaluated at the predicted position. Completing
the square in the Gaussian log posterior gives $\mathbf J_m^t$ as the
precision and $\mathbf z_m^t$ as the linear term in the position
increment. In particular,
$\mathbf P_m^{t|t}\preceq\mathbf\Sigma_m^{t|t-1}$ since the added
information is positive semidefinite. With zero new sensing power,
the posterior equals the prior. The quality indicator reported in
\eqref{eq:sensing_crlb} is $\sqrt{\operatorname{tr}\mathbf P_m^{t|t}}$
under the same local model. The sensing report is
\begin{equation}
	\mathcal O_k^{s,t}
	=\{(m,\widehat{\mathbf q}_m^t,\mathrm{CRLB}_m^t):
	m\in\mathcal M_k^t\}.
	\label{eq:sensing_observation}
\end{equation}
The BS shares the same estimate and quality indicator with all agents
requesting target $m$.

To prepare the next decision, the BS propagates each posterior through
\eqref{eq:target_dynamics}. Let $\mathbf Q_m^t$ be the covariance of the
independent, zero-mean process noise and $\mathbf F_m^t$ the Jacobian of
$\mathcal F_m$ at $\widehat{\mathbf q}_m^t$. The mean and covariance of the first-order Gaussian predictive
distribution are
\begin{align}
	\widehat{\mathbf q}_m^{t+1|t}
	&=\mathcal F_m(\widehat{\mathbf q}_m^t),\notag\\
	\mathbf\Sigma_m^{t+1|t}
	&=\mathbf F_m^t\mathbf P_m^{t|t}(\mathbf F_m^t)^T+\mathbf Q_m^t.
	\label{eq:feedback_prediction}
\end{align}
The Gaussian prediction is exact for affine dynamics and Gaussian
process noise under the adopted posterior model. For affine dynamics,
propagation preserves the covariance ordering between the predictive
distributions obtained with and without new measurements from the same prior.
Process noise can still increase uncertainty between epochs. We assume
the predicted covariance is positive definite as in
\eqref{eq:sensing_prediction_prior}, which holds if $\mathbf Q_m^t\succ\mathbf 0$.
At $\tau^{t+1}$, the BS distributes a control-plane map
$\mathcal B^{t+1}$ containing the predictive position
distributions, represented by their means and covariances, along with
IDs and class labels for all initialized tracks. Each agent sets
$\mathbf b_k^{t+1}=\mathcal B^{t+1}$. Tracks outside $\mathcal M_k^t$
can thus enter future requests as their predicted positions change.
Map signaling is outside the modeled resource budget.

\textit{3) Regenerate the request.}
At the epoch boundary, the agent receives the feedback
$\mathbf o_k^t=(\mathcal O_k^{c,t},\mathcal O_k^{s,t},\mathcal B^{t+1})$.
Its task state is
$\mathbf x_k^{t+1}=(\mathcal Q_k^{t+1,-},\mathcal E_k^{t+1},
\mathcal U_k^{t+1})$, where $\mathcal E_k^{t+1}$ and $\mathcal U_k^{t+1}$
are the newly available events and the available data products,
respectively. The application can use completed payloads and the
predictive distributions in $\mathbf b_k^{t+1}$ to generate these inputs. For an intent with an
obstacle-refinement clause, let
$\mathcal C_{k,\Omega}^{t+1}$ denote the obstacle tracks
predicted to lie in a task-relevant region $\Omega$. An application
can generate a refinement event for the next decision whenever
\begin{equation}
	\exists\,m\in\mathcal C_{k,\Omega}^{t+1}:
	\sqrt{\operatorname{tr}\mathbf\Sigma_m^{t+1|t}}
	>\epsilon_k^{\mathrm{ref}},
	\label{eq:feedback_refinement_trigger}
\end{equation}
where $\epsilon_k^{\mathrm{ref}}$ is the task's localization tolerance.
The application emits one refinement event per region at each epoch
boundary where the condition holds. The event activates a clause with
target limit one, relevance score
$\sqrt{\operatorname{tr}\mathbf\Sigma_m^{t+1|t}}$, and tolerance
$\epsilon_k^{\mathrm{ref}}$, requesting the most uncertain obstacle
in $\Omega$. The extra request ends when the condition clears and can
resume if motion or process noise raises prediction uncertainty again.
The agent then evaluates the clauses with the same user intent,
\begin{equation}
	\mathbf g_k^{t+1}
	=\mathcal G_k(\mathcal I_k,\mathbf x_k^{t+1},\mathbf b_k^{t+1}).
	\label{eq:feedback_next_request}
\end{equation}
If neither service is needed, the agent skips request submission.
Thus, packet delivery and expiration change the communication demand,
while the updated predictive distributions change sensing targets, relevance,
and the activation of refinement clauses. The BS allocates resources
in the next epoch using these updated requests.

\subsection{Value of Service Model}
We first define a normalized evaluation metric
for the services provided under this goal, and then formulate
the corresponding resource allocation problem. The metric
evaluates goal satisfaction at epoch $t$ according to the user intent.

For agent $k$, the communication requirement is
$R_{k,\mathrm{req}}^t=D_k^t/L_k^t$, where $D_k^t\geq0$
and $L_k^t>0$. We define the communication utility as \cite{10136792}
\begin{align}
	S_k^{c,t}
	=
	\begin{cases}
		\displaystyle
		\min\left\{
		\frac{R_k^t}{R_{k,\mathrm{req}}^t},1
		\right\},
		& D_k^t>0,\\[1ex]
		1,
		& D_k^t=0.
	\end{cases}
	\label{eq:communication_fulfillment}
\end{align}
Here, $R_k^t$ is the achievable rate in \eqref{eq:communication_rate}.
The score measures how well the current allocation meets the rate
implied by the requested data volume and delivery budget. For $D_k^t>0$,
$S_k^{c,t}=1$ means that $R_k^t$ meets or exceeds $D_k^t/L_k^t$.
Actual delivery during epoch $t$ depends on its duration and the packet
deadlines and is accounted for separately in
\eqref{eq:feedback_packet_service}--\eqref{eq:feedback_delivered_bits}.

For each requested target $m\in\mathcal{M}_k^t$, we define \cite{10136792}
the sensing utility as 
\begin{align}
	S_{k,m}^{s,t}
	=
	\min\left\{
	\frac{\epsilon_{k,m}^t}{\mathrm{CRLB}_m^t},1
	\right\},
	\qquad m\in\mathcal{M}_k^t,
	\label{eq:target_sensing_fulfillment}
\end{align}
where $\epsilon_{k,m}^t>0$ is the tolerable localization error
specified by the agent and $\mathrm{CRLB}_m^t$ is the error-bound
indicator derived in \eqref{eq:sensing_crlb}. For evaluation, we set
$\mathrm{CRLB}_m^t=+\infty$ when the information matrix is
singular, yielding $S_{k,m}^{s,t}=0$. The score is one
in the limiting case $\mathrm{CRLB}_m^t=0$.

For all sensing targets requested by agent $k$, the sensing utility is calculated with
\begin{align}
	S_k^{s,t}
	=
	\sum_{m\in\mathcal{M}_k^t}
	\phi_{k,m}^t S_{k,m}^{s,t},
	\label{eq:sensing_fulfillment}
\end{align}
where the nonnegative relevance weights satisfy
$\sum_{m\in\mathcal{M}_k^t}\phi_{k,m}^t=1$ whenever
$\mathcal{M}_k^t\neq\emptyset$. When
$\mathcal{M}_k^t=\emptyset$, we set $S_k^{s,t}=0$.

Let $\mathcal K_{\mathrm{act}}^t$ be the set of agents that submit a
request at epoch $t$. For each $k\in\mathcal K_{\mathrm{act}}^t$, the
value of service (VoS) is \cite{10136792}
\begin{align}
	V_k^{g,t}
	=
	\alpha_k^t S_k^{c,t}
	+(1-\alpha_k^t)S_k^{s,t}.
	\label{eq:goal_value}
\end{align}
Thus, $V_k^{g,t}\in[0,1]$ combines communication and sensing
satisfaction according to their importance to the current
task. It is a soft evaluation metric rather than a set of
mandatory service guarantees.

Let $\omega_k>0$, with $\sum_{k=1}^{K}\omega_k=1$, denote the fixed
system-level priority of agent $k$. Set $V_k^{g,t}=0$ for
$k\notin\mathcal K_{\mathrm{act}}^t$. The weights retain their original
values as the active set changes, giving
\begin{align}
	V_{\mathrm{sys}}^{g,t}
	=
	\sum_{k\in\mathcal K_{\mathrm{act}}^t}\omega_k V_k^{g,t}.
	\label{eq:system_goal_value}
\end{align}
When $\mathcal K_{\mathrm{act}}^t=\emptyset$, the system VoS is zero
and the BS leaves the resources idle.
When several agents request the same target, its common error-bound indicator enters
their respective satisfaction scores, each evaluated using
that agent's own accuracy requirement and relevance weight.

\subsection{Problem Formulation}

At the beginning of decision epoch $t$, the BS receives the
current structured goals $\{\mathbf{g}_k^t:k\in\mathcal K_{\mathrm{act}}^t\}$.
Given these goals, the current channel information, and the
propagated sensing priors, the BS selects
$\mathbf{R}^t
=\{\boldsymbol{\chi}^{c,t},\boldsymbol{\chi}^{s,t},
\mathbf{p}^{c,t},\mathbf{p}^{s,t}\}$
to maximize the current VoS of system. The transmit
beamformers are treated as given during this optimization.
For each $k\notin\mathcal K_{\mathrm{act}}^t$, we fix
$\chi_{k,n}^{c,t}=p_{k,n}^{c,t}=0$ on all subchannels.
The resulting problem is
\begin{subequations}
	\label{prob:P1}
	\begin{align}
		\mathcal{P}_1^t:
		\max_{\mathbf{R}^t}\,
		&
		\sum_{k\in\mathcal K_{\mathrm{act}}^t}\omega_k \! \!
		\left[
		\alpha_k^t S_k^{c,t}
		\! + \!\! (1-\alpha_k^t) \!
		\! \! \sum_{m\in\mathcal{M}_k^t} \! \!
		\phi_{k,m}^t S_{k,m}^{s,t}
		\right]
		\tag{\theparentequation}
		\label{prob:P1_obj}
		\\
		\mathrm{s.t.}\quad
		&
		\sum_{k=1}^{K}\chi_{k,n}^{c,t}
		+\sum_{m=1}^{M}\chi_{m,n}^{s,t}
		\leq 1,
		\quad \forall n\in\mathcal{N},
		\label{prob:P1_subchannel}
		\\
		&
		\sum_{n=1}^{N}
		\left(
		\sum_{k=1}^{K}\chi_{k,n}^{c,t}p_{k,n}^{c,t}
		+
		\sum_{m=1}^{M}\chi_{m,n}^{s,t}p_{m,n}^{s,t}
		\right)
		\leq P_{\max},
		\label{prob:P1_power}
		\\
		&
		\chi_{k,n}^{c,t}\in\{0,1\},
		\quad \forall k,n,
		\label{prob:P1_binary_c}
		\\
		&
		\chi_{m,n}^{s,t}\in\{0,1\},
		\quad \forall m,n,
		\label{prob:P1_binary_s}
		\\
		&
		p_{k,n}^{c,t}\geq 0,
		\quad \forall k,n,
		\label{prob:P1_nonnegative_c}
		\\
		&
		p_{m,n}^{s,t}\geq 0,
		\quad \forall m,n.
		\label{prob:P1_nonnegative_s}
	\end{align}
\end{subequations}

Constraint \eqref{prob:P1_subchannel} ensures that each subchannel
is assigned to at most one communication user or one sensing
target at decision epoch $t$. Constraint \eqref{prob:P1_power}
limits the total transmit power allocated to communication and
sensing operations to the BS power budget $P_{\max}$.
Constraints \eqref{prob:P1_binary_c} and
\eqref{prob:P1_binary_s} specify the binary communication and
sensing subchannel assignments, respectively, where a value
of one indicates assignment and zero indicates no assignment.
Finally, constraints \eqref{prob:P1_nonnegative_c} and
\eqref{prob:P1_nonnegative_s} ensure that the communication
and sensing transmit powers are  nonnegative.


\section{Methodology}
\label{sec:methodology}
Problem $\mathcal{P}_1^t$ couples the agents' service requirements
through shared subchannel and transmit-power budgets. Communication
directly serves one agent, while sensing a target can improve the
service values of all agents requesting it.

We develop a welfare-oriented Stackelberg mechanism in which
the BS acts as the leader and the agents act as followers.
The BS offers subchannel access, power allowances, and
personalized virtual prices, while each agent determines its
communication and sensing contributions.
The prices account for the benefits that a sensing contribution
provides to other agents, enabling a system-optimal allocation
to be supported as a follower equilibrium.
The BS maximizes system VoS rather than pricing revenue.

\subsection{Stackelberg Interaction and Shared-Sensing Contributions}
\label{subsec:stackelberg_interaction}

At the beginning of epoch $t$, agent $k$ generates
$\mathbf{g}_k^t
=\mathcal{G}_k(\mathcal{I}_k,\mathbf{x}_k^t,\mathbf{b}_k^t)$.
The mechanism takes the structured goals as inputs. The goals,
channel information, sensing priors, and beamformers are fixed during
allocation. We omit the epoch superscript $t$ where clear.

The BS selects the subchannel configuration
$\boldsymbol{\chi}
=\{\boldsymbol{\chi}^{c},\boldsymbol{\chi}^{s}\}$.
Let $\mathcal{X}$ denote the configurations satisfying the
binary assignment and subchannel exclusivity constraints
of $\mathcal{P}_1$.
Targets outside $\bigcup_k\mathcal{M}_k$ contribute zero to the
current objective, so excluding them preserves the optimal value
of $\mathcal{P}_1$.

Define the offered communication and sensing subchannels as
$\mathcal{N}_k^c=\{n:\chi_{k,n}^c=1\}$ and
$\mathcal{N}_m^s=\{n:\chi_{m,n}^s=1\}$, respectively.
Each offered subchannel has bandwidth $\Delta f$.
The BS also assigns nonnegative power allowances satisfying
$\sum_k\bar P_k\leq P_{\max}$.
An allowance authorizes the use of BS transmit power;
it does not represent an agent-side transmission or require
the agent to consume its entire allowance.

Agent $k$ chooses communication contributions $z_{k,n}^{c}$
and sensing contributions $z_{k,m,n}^{s}$ over its offered
operations.
Its contribution vector $\mathbf{z}_k$ belongs to
\begin{align}
	\mathcal{Z}_k(\boldsymbol{\chi},\bar P_k)
	=&
	\Bigg\{
	\mathbf{z}_k\geq\mathbf{0}:\;
	\sum_{n\in\mathcal{N}_k^c}z_{k,n}^{c}
	\nonumber\\
	&+
	\sum_{m\in\mathcal{M}_k}
	\sum_{n\in\mathcal{N}_m^s}z_{k,m,n}^{s}
	\leq\bar P_k
	\Bigg\}.
	\label{eq:sg_agent_strategy}
\end{align}
An agent with no offered operations has the singleton
strategy set $\{\mathbf{0}\}$.

Let $\mathcal{K}_m=\{k:m\in\mathcal{M}_k\}$ denote the agents
requesting target $m$.
The BS converts the contributions into physical powers as
\begin{align}
	p_{k,n}^{c}
	&=z_{k,n}^{c},
	&& n\in\mathcal{N}_k^c,
	\nonumber\\
	p_{m,n}^{s}
	&=\sum_{k\in\mathcal{K}_m}z_{k,m,n}^{s},
	&& n\in\mathcal{N}_m^s.
	\label{eq:sg_physical_supply}
\end{align}
Unassigned powers are set to zero.
The sensing contributions determine the power of each shared
target-level operation.
All agents requesting target $m$ therefore use the same
$\mathrm{CRLB}_m$, evaluated against their respective
accuracy requirements and relevance weights.

Each contribution is counted once in the total physical power.
The total is at most $\sum_k\bar P_k\leq P_{\max}$, so every admissible
follower strategy profile is physically feasible.

The BS announces nonnegative prices
$\pi_{k,n}^{c}$ and $\pi_{k,m,n}^{s}$ per unit contributed
power.
Agent $k$ incurs the virtual payment
\begin{align}
	T_k
	=
	&\sum_{n\in\mathcal{N}_k^c}
	\pi_{k,n}^{c}z_{k,n}^{c}
	+
	\sum_{m\in\mathcal{M}_k}
	\sum_{n\in\mathcal{N}_m^s}
	\pi_{k,m,n}^{s}z_{k,m,n}^{s}.
	\label{eq:sg_contribution_payment}
\end{align}
These virtual payments are expressed in the same units as weighted
VoS and serve as follower incentives.

Let
$\boldsymbol{\theta}
=(\boldsymbol{\chi},\{\bar P_k\},\{\boldsymbol{\pi}_k\})$
collect the leader's decisions.
The configured follower utility is
\begin{align}
	U_k(\mathbf{z}_k,\mathbf{z}_{-k};\boldsymbol{\theta})
	=
	\omega_k
	V_k^g\bigl(\boldsymbol{\chi},\mathbf{p}(\mathbf{z})\bigr)
	-T_k.
	\label{eq:sg_actual_utility}
\end{align}
An agent benefits from the total sensing power contributed
to its requested targets while paying for its own contribution.
The system-prescribed weights $\omega_k$ form part of the
configured utilities.

\begin{definition}[Follower equilibrium]
	\label{def:sg_follower_equilibrium}
	For a fixed leader decision $\boldsymbol{\theta}$,
	a feasible contribution profile $\mathbf{z}^{\star}$
	is a follower Nash equilibrium if
	\begin{align}
		U_k(\mathbf{z}_k^{\star},\mathbf{z}_{-k}^{\star};
		\boldsymbol{\theta})
		\geq
		U_k(\widetilde{\mathbf{z}}_k,\mathbf{z}_{-k}^{\star};
		\boldsymbol{\theta})
		\label{eq:sg_follower_ne}
	\end{align}
	for every agent $k$ and every
	$\widetilde{\mathbf{z}}_k\in
	\mathcal{Z}_k(\boldsymbol{\chi},\bar P_k)$.
	The set of such equilibria is denoted by
	$\mathcal{E}(\boldsymbol{\theta})$.
\end{definition}

Prices belong to an admissible box
$\Pi=\{\boldsymbol{\pi}:0\leq\pi_i\leq\pi_{\max}\}$,
where $i$ indexes the offered communication and sensing
price entries.
The bound $\pi_{\max}$ is chosen to contain the supporting
prices derived below.
Let $\Theta$ denote all leader decisions satisfying
$\boldsymbol{\chi}\in\mathcal{X}$,
$\boldsymbol{\pi}\in\Pi$,
$\bar P_k\geq0$, and
$\sum_k\bar P_k\leq P_{\max}$.

We assume optimistic selection. Among the equilibria in
$\mathcal{E}(\boldsymbol{\theta})$, the selected response maximizes
the BS's VoS. This selection rule is part of the mechanism.

\begin{definition}[Optimistic Stackelberg--Nash equilibrium]
	\label{def:sg_stackelberg_equilibrium}
	A pair $(\boldsymbol{\theta}^{\star},\mathbf{z}^{\star})$
	is an optimistic Stackelberg--Nash equilibrium if it is a
	global solution of
	\begin{align}
		\mathcal{P}_{\mathrm{SG}}:\quad
		\max_{\boldsymbol{\theta},\mathbf{z}}\quad&
		V_{\mathrm{sys}}^g
		\bigl(\boldsymbol{\chi},\mathbf{p}(\mathbf{z})\bigr)
		\nonumber\\
		\mathrm{s.t.}\quad&
		\boldsymbol{\theta}\in\Theta,\qquad
		\mathbf{z}\in\mathcal{E}(\boldsymbol{\theta}).
		\label{prob:sg_leader}
	\end{align}
\end{definition}

The BS objective is system VoS. The analysis takes reported goals,
the prescribed follower utilities, and the selection rule as given.
Truthful reporting and monetary cost-sharing fairness are outside
its scope.

\subsection{Supporting Prices and Equilibrium Existence}
\label{subsec:sg_existence}

The analysis uses the sensing model in Section~II with
$\mathbf{J}_{m,n}(p)=p\mathbf{H}_{m,n}$,
$\mathbf{H}_{m,n}\succeq\mathbf{0}$, and
$\mathbf{J}_{m,0}\succ\mathbf{0}$.
The information matrices and sensing parameters remain
fixed during power optimization.
We also assume $P_{\max}>0$, finite channel gains,
positive communication noise power, and
$R_{k,\mathrm{req}}>0$ whenever $D_k>0$.

\begin{lemma}[Concavity of service value]
	\label{lem:sg_concavity}
	For a fixed subchannel configuration
	$\boldsymbol{\chi}$, each
	$F_k(\mathbf{p})
	=\omega_kV_k^g(\boldsymbol{\chi},\mathbf{p})$
	is continuous, concave, and nondecreasing in the assigned
	physical powers.
\end{lemma}

\textit{Proof:}
	The communication rate is a sum of concave logarithmic
	functions of power.
	For $D_k>0$, composing this rate with the concave,
	nondecreasing function
	$r\mapsto\min\{r/R_{k,\mathrm{req}},1\}$
	preserves concavity.
	For $D_k=0$, the communication score is constant.
	
	For sensing, define
	$h(\mathbf{J})=1/\operatorname{tr}(\mathbf{J}^{-1})$
	on the real symmetric positive-definite matrices.
	It admits the representation
	\begin{align}
		h(\mathbf{J})
		=
		\min_{\substack{\mathbf{Z}\in\mathbb{R}^{2\times2}\\
				\operatorname{tr}(\mathbf{Z})=1}}
		\operatorname{tr}(\mathbf{Z}^{T}\mathbf{J}\mathbf{Z}).
		\label{eq:sg_information_variational}
	\end{align}
	The minimum is attained at
	$\mathbf{Z}
	=\mathbf{J}^{-1}/\operatorname{tr}(\mathbf{J}^{-1})$.
	Since the minimized expression is affine in $\mathbf{J}$,
	$h$ is concave.
	
	The accumulated information matrix is affine in power.
	Consequently, the target satisfaction score
	$\min\{\epsilon_{k,m}\sqrt{h(\mathbf{J}_m)},1\}$
	is concave, because square root and saturation at one
	are concave nondecreasing operations.
	Increasing sensing power adds positive-semidefinite
	information and cannot decrease the score.
	
	The nonnegative weights $\alpha_k$, $1-\alpha_k$,
	$\phi_{k,m}$, and $\omega_k$ preserve these properties.
	Positive-definite priors ensure continuity over all
	feasible powers, including zero sensing power.

\begin{lemma}[Existence of follower equilibrium]
	\label{lem:sg_follower_existence}
	For every admissible leader decision
	$\boldsymbol{\theta}\in\Theta$,
	the contribution game has a pure-strategy Nash equilibrium.
\end{lemma}

\textit{Proof:}
Each $\mathcal{Z}_k$ is nonempty, compact, and convex.
For fixed $\mathbf{z}_{-k}$, the physical powers are affine
in $\mathbf{z}_k$.
Lemma~\ref{lem:sg_concavity} and the linear payment therefore
make $U_k$ concave in its own strategy and continuous
in all strategies.

Each best-response correspondence has nonempty, compact,
convex values and is upper hemicontinuous.
Their product maps the compact convex joint strategy set
into itself.
Kakutani's fixed-point theorem yields a fixed point,
which satisfies Definition~\ref{def:sg_follower_equilibrium}.

We next construct prices that support a service-optimal
physical allocation.
For fixed $\boldsymbol{\chi}$, define
\begin{align}
	W(\boldsymbol{\chi})
	=
	\max_{\mathbf{p}\in\mathcal{P}(\boldsymbol{\chi})}
	\sum_{k=1}^{K}F_k(\mathbf{p}),
	\label{eq:sg_fixed_configuration_value}
\end{align}
where $\mathcal{P}(\boldsymbol{\chi})$ contains the nonnegative
assigned powers satisfying the BS total-power constraint;
unassigned powers are fixed to zero.
By Lemma~\ref{lem:sg_concavity}, this is a convex
optimization problem in maximization form.

For a configuration with at least one offered operation,
$P_{\max}>0$ provides a strictly feasible power vector,
so the KKT conditions apply.
Let $\mathbf{p}^{\star}$ be an optimum and
$\lambda_P\geq0$ its total-power multiplier.
Choose nonnegative supergradients
$\boldsymbol{\mu}_k\in\partial^{+}F_k(\mathbf{p}^{\star})$
that jointly satisfy the KKT conditions.
Here, a supergradient satisfies
$F_k(\mathbf{p})
\leq F_k(\mathbf{p}^{\star})
+\boldsymbol{\mu}_k^{T}(\mathbf{p}-\mathbf{p}^{\star})$.

Write $\mu_{k,n}^{c}$ for the marginal value of agent $k$'s
communication power and $\mu_{j,m,n}^{s}$ for the marginal
value that agent $j$ receives from a sensing operation.
Set $\mu_{j,m,n}^{s}=0$ when $m\notin\mathcal{M}_j$.
At differentiable points these quantities are partial
derivatives; at saturation points they are compatible
supergradients, rather than independently chosen derivatives.

For each offered operation, the KKT conditions give
\begin{gather}
	\mu_{k,n}^{c} \leq \lambda_P, \nonumber\\
	p_{k,n}^{c,\star}\bigl(\lambda_P-\mu_{k,n}^{c}\bigr)=0, \nonumber\\
	\sum_j \mu_{j,m,n}^{s} \leq \lambda_P, \nonumber\\
	p_{m,n}^{s,\star}\Bigl(\lambda_P-\sum_j \mu_{j,m,n}^{s}\Bigr)=0,
	\nonumber\\
	\lambda_P\Bigl(P_{\max}
	-\sum_{k,n} p_{k,n}^{c,\star}
	-\sum_{m,n} p_{m,n}^{s,\star}\Bigr)=0.
	\label{eq:sg_social_marginal_condition}
\end{gather}
The power sums run over the assigned operations.

The prices can be constructed as
\begin{align}
	\pi_{k,n}^{c}
	&=\lambda_P,
	\nonumber\\
	\pi_{k,m,n}^{s}
	&=\lambda_P-\sum_{j\neq k}\mu_{j,m,n}^{s}.
	\label{eq:sg_supporting_prices}
\end{align}
For sensing, the price deducts the marginal benefits
received by other agents from the resource scarcity price.
At $\mathbf{p}^{\star}$, this gives
\begin{align}
	\mu_{k,m,n}^{s}-\pi_{k,m,n}^{s}
	=
	\sum_j\mu_{j,m,n}^{s}-\lambda_P.
	\label{eq:sg_marginal_alignment}
\end{align}
At the supported allocation $\mathbf{p}^{\star}$, the agent's net
marginal utility equals the system's net marginal service value.
The announced prices remain fixed during unilateral deviations.

The prices are nonnegative by
\eqref{eq:sg_social_marginal_condition}.
For each epoch, the uncapped satisfaction functions have
finite derivatives over the compact power domain:
the communication denominators are positive, and the sensing
information matrices remain positive definite.
The finite configuration set therefore admits a common
finite upper bound on the selected system marginal values.
A supporting multiplier can be chosen within this bound,
and hence so can the prices.
We choose $\pi_{\max}$ at least as large as this bound.
For the configuration with no offered operations, take
$\lambda_P=0$.

To implement $\mathbf{p}^{\star}$, choose nonnegative
contributions satisfying
$z_{k,n}^{c,\star}=p_{k,n}^{c,\star}$ and
$\sum_{k\in\mathcal{K}_m}z_{k,m,n}^{s,\star}
=p_{m,n}^{s,\star}$.
For example, divide each sensing power equally among
its requesting agents.
Set
\begin{align}
	\bar P_k
	=
	\sum_{n\in\mathcal{N}_k^c}z_{k,n}^{c,\star}
	+
	\sum_{m\in\mathcal{M}_k}
	\sum_{n\in\mathcal{N}_m^s}z_{k,m,n}^{s,\star}.
	\label{eq:sg_supporting_allowances}
\end{align}
The selected allocation determines the allowances. Equal splitting
is one feasible way to implement the sensing contributions.

\begin{proposition}
	\label{prop:sg_implementation}
	For any fixed configuration, the optimal physical allocation
	$\mathbf{p}^{\star}$ can be implemented as a follower Nash
	equilibrium using
	\eqref{eq:sg_supporting_prices} and
	\eqref{eq:sg_supporting_allowances}.
\end{proposition}

\textit{Proof:}
The constructed profile is feasible and implements
$\mathbf{p}^{\star}$.
Let $\mathbf{h}_k$ be the supergradient of agent $k$'s
utility with respect to its own contributions at this profile.
For communication its entries are
$\mu_{k,n}^{c}-\lambda_P$.
For sensing they are given by
\eqref{eq:sg_marginal_alignment}.

The KKT conditions imply $\mathbf{h}_k\leq\mathbf{0}$.
Moreover, every positive contribution has a zero
corresponding entry of $\mathbf{h}_k$, since a positive
contribution implies positive physical power.
Thus, $\mathbf{h}_k^{T}\mathbf{z}_k^{\star}=0$.

For any feasible unilateral alternative
$\widetilde{\mathbf{z}}_k$, concavity yields
\begin{align}
	&U_k(\widetilde{\mathbf{z}}_k,\mathbf{z}_{-k}^{\star};
	\boldsymbol{\theta})
	-U_k(\mathbf{z}^{\star};\boldsymbol{\theta})
	\nonumber\\
	&\quad\leq
	\mathbf{h}_k^{T}
	(\widetilde{\mathbf{z}}_k-\mathbf{z}_k^{\star})
	=
	\mathbf{h}_k^{T}\widetilde{\mathbf{z}}_k
	\leq0.
	\label{eq:sg_implementation_proof}
\end{align}
No agent can improve its utility unilaterally.

\begin{proposition}[Existence and optimal-value equivalence]
	\label{prop:sg_global_equilibrium}
	If $\Pi$ contains the supporting prices, an optimistic
	Stackelberg--Nash equilibrium exists, and
	$\mathcal{P}_{\mathrm{SG}}$ has the same optimal value
	as $\mathcal{P}_1$.
\end{proposition}

\textit{Proof:}
After setting unassigned powers to zero,
$\mathcal{P}_1$ has finitely many configurations, each with
a continuous objective over a compact power domain.
It therefore has a global optimum
$(\boldsymbol{\chi}^{\star},\mathbf{p}^{\star})$.

Proposition~\ref{prop:sg_implementation} constructs an
admissible leader decision and a follower equilibrium
implementing this optimum.
Conversely, \eqref{eq:sg_physical_supply} and the allowance
constraints make every admissible game outcome feasible
for $\mathcal{P}_1$.
No such outcome can exceed its global optimal value.

The constructed equilibrium attains that value.
Under optimistic selection it is therefore an admissible
selected response, and no alternative leader decision can
achieve a higher VoS. These results establish equilibrium existence and
implementation of the centralized optimum.
They do not establish uniqueness or convergence of arbitrary
follower best-response dynamics.
In particular, the optimistic selection assumption remains
necessary for the stated leader-level result.

\subsection{Online Allocation and Agentic Feedback}
\label{subsec:sg_computation}

For each candidate configuration, the BS solves
\eqref{eq:sg_fixed_configuration_value} using an interior-point or
another convergent convex solver. It uses the power multiplier and
compatible marginal values to construct prices, contribution shares,
and allowances.

Prices follow from the computed allocation. The solver iterations
optimize physical power, and the follower equilibrium is checked
for the resulting offer.
The recommended contribution profile is announced with the
leader's offer.
For an exact solution, Proposition~\ref{prop:sg_implementation}
ensures that following the recommendation is individually
optimal when the other agents do so.
Its adoption remains subject to the stated selection protocol.

For numerical solutions, define the unilateral deviation gain
\begin{align}
	\delta_k
	=
	\max_{\widetilde{\mathbf{z}}_k\in\mathcal{Z}_k}
	U_k(\widetilde{\mathbf{z}}_k,\mathbf{z}_{-k};
	\boldsymbol{\theta})
	-
	U_k(\mathbf{z};\boldsymbol{\theta}).
	\label{eq:sg_follower_gap}
\end{align}
A profile with $\max_k\delta_k\leq\tau_F$ is a $\tau_F$-Nash
equilibrium. Each agent can gain at most $\tau_F$ by deviating
unilaterally.

Each deviation problem is a convex optimization problem
in maximization form.
In numerical implementation, let $\widehat{\delta}_k$
be a certified upper bound on $\delta_k$, obtained from
an upper bound on the deviation problem's optimal value.
A feasible deviation gives a lower bound on the possible gain.
Certification requires an upper bound.
A candidate passes the follower check when
$\max_k\widehat{\delta}_k\leq\tau_F$.

The candidate powers and allowances must also be physically
feasible, and its prices must belong to $\Pi$.
Prices computed from approximate multipliers may be projected
onto $\Pi$ before the follower check.
This numerical correction does not replace the exact
supporting-price result; its effect is assessed by the
deviation-gap certificate.
A candidate failing the checks is refined within the
numerical budget or discarded.

Exhaustively evaluating every configuration with exact
continuous solutions recovers the equilibrium characterized
in Proposition~\ref{prop:sg_global_equilibrium}.
\textcolor{black}{For online operation, we use the neighborhood
	$\mathcal{H}(\boldsymbol{\chi})$ consisting of feasible
	single-subchannel reassignments and pairwise assignment swaps.}
Reassignments may include the idle state.

\textcolor{black}{A cold start uses the all-idle configuration, which provides a
	feasible zero-contribution follower equilibrium. For a warm start,
	the BS reuses the preceding epoch's assignment and sets subchannels
	assigned to targets outside $\bigcup_k\mathcal{M}_k$ to idle. It solves
	\eqref{eq:sg_fixed_configuration_value} for the resulting assignment
	under the current goals and recomputes the prices, contributions, and
	allowances. This solve counts toward the candidate-solve budget. The
	BS adopts the warm-start allocation as the incumbent if it passes
	the feasibility and follower checks. If no prior assignment is
	available or certification fails within the budget, the BS uses the
	all-idle configuration.}
At each iteration, the BS evaluates neighboring configurations
and accepts the best certified candidate only if its achieved
VoS exceeds the incumbent value by more than
$\tau_{\mathrm{out}}>0$.
The incumbent is retained whenever no candidate is accepted
or the computation budget is exhausted.

\begin{algorithm}[!b]
	\caption{Supporting-Price Resource Allocation}
	\label{alg:sg_online}
	\begin{algorithmic}[1]
		
		\REQUIRE \textcolor{black}{$P_{\max}$, $\Pi$, $\tau_F,\tau_{\mathrm{out}}>0$,
			computation budgets, and initialization mode}
		
		\FOR{each epoch $t$}
		\STATE Receive goals $\{\mathbf{g}_k^t\}$ and network information
		\STATE \textcolor{black}{Initialize the incumbent and its VoS $v$ using the
			all-idle allocation for a cold start or the screened previous
			assignment for a warm start (Section~\ref{subsec:sg_computation}).}
		
		\WHILE{search budget remains}
		\STATE Set $\mathcal{A}\gets\emptyset$
		
		\FOR{each neighbor
			$\boldsymbol{\chi}'\in\mathcal{H}(\boldsymbol{\chi})$
			within budget}
		\STATE Solve \eqref{eq:sg_fixed_configuration_value};
		construct prices, contributions, and allowances
		\STATE Add the candidate to $\mathcal{A}$ if feasible,
		prices belong to $\Pi$, and
		$\max_k\widehat{\delta}_k\leq\tau_F$
		\ENDFOR
		
		\IF{$\mathcal{A}=\emptyset$}
		\STATE Exit the search loop
		\ELSE
		\STATE Select the best candidate with VoS $v'$
		\IF{$v'>v+\tau_{\mathrm{out}}$}
		\STATE Update the incumbent; $v\gets v'$
		\ELSE
		\STATE Exit the search loop
		\ENDIF
		\ENDIF
		\ENDWHILE
		
		\STATE Announce the incumbent offer and recommended contributions
		\STATE Execute $\mathbf{R}^t$ under the selection protocol
		\STATE Return $\{\mathbf{o}_k^t\}$ and update agent states
		\ENDFOR
		
	\end{algorithmic}
\end{algorithm}

Since $V_{\mathrm{sys}}^g\in[0,1]$, the search can accept only
finitely many gains exceeding $\tau_{\mathrm{out}}$.
\textcolor{black}{With exact inner solutions and a complete final neighborhood
	evaluation, termination means that no configuration in
	$\mathcal{H}(\boldsymbol{\chi})$ improves
	$W(\boldsymbol{\chi})$ by more than $\tau_{\mathrm{out}}$.}
Numerical candidate failures or termination at a computation
limit do not provide this exact neighborhood guarantee.

For a returned feasible allocation, a valid upper bound
$V_{\mathrm{UB}}$ on $\mathcal{P}_1$ gives the certificate
\begin{align}
	V_{\mathrm{UB}}
	-
	V_{\mathrm{sys}}^g(\boldsymbol{\chi},\mathbf{p})
	\leq\tau_L
	\label{eq:sg_global_value_gap}
\end{align}
bounds its loss relative to the global optimum by $\tau_L$.
The normalization gives the valid, potentially loose,
bound $V_{\mathrm{UB}}=1$.
Tighter bounds can be obtained through a globally valid
relaxation or exhaustive evaluation in small instances.

The follower gap and global value gap certify different
properties: the former bounds unilateral follower gains,
while the latter bounds the loss of the implemented physical
allocation.
They do not, by themselves, prove that a numerically returned
profile is the exact optimistic response to its leader decision. 
We therefore describe the online procedure as an approximate
solution method for the Stackelberg formulation, rather than
claiming that every returned solution is a global
Stackelberg--Nash equilibrium.

After allocation, the BS executes
$\mathbf{R}^t
=\{\boldsymbol{\chi}^{c,t},\boldsymbol{\chi}^{s,t},
\mathbf{p}^{c,t},\mathbf{p}^{s,t}\}$.
The feedback $\mathbf{o}_k^t$ updates the task state
$\mathbf{x}_k^{t+1}$ and the predictive distributions in
$\mathbf{b}_k^{t+1}$ as in Section~II. The agent then generates
$\mathbf{g}_k^{t+1}
=\mathcal{G}_k(\mathcal{I}_k,\mathbf{x}_k^{t+1},
\mathbf{b}_k^{t+1})$.
The allocation is recomputed with the updated goals and
network information, while the original intent remains fixed.
Algorithm~\ref{alg:sg_online} summarizes the procedure.

\section{Simulation}

\begin{figure*}[htbp]
	\centering
	\includegraphics[width=\textwidth]{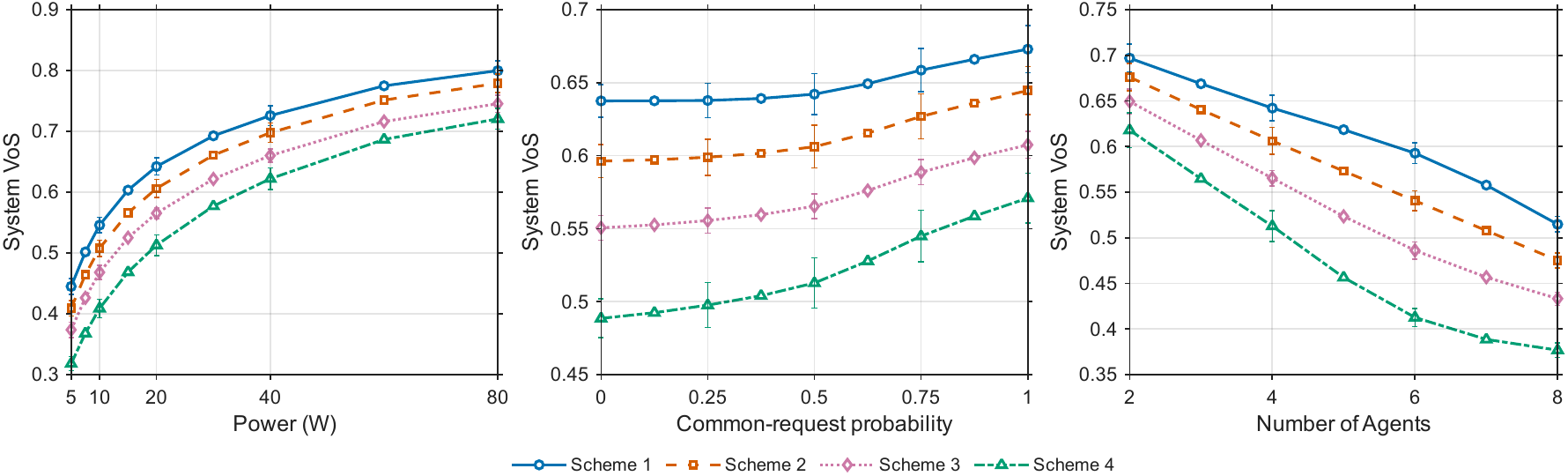}
	\caption{System VoS versus (a) the transmit power
		budget, (b) the common-request probability, and (c) the number of
		agents. Error bars are shown at the labeled sweep points.
		Intermediate markers are interpolated.}
	\label{fig:sim_resources}
\end{figure*}

\begin{table*}[b]
	\centering
	\caption{Closed-loop multi-epoch performance across Schemes~1--4 and the Oracle reference (see Fig.~\ref{fig:sim_feedback}). Values report the mean ($\pm$ standard error) of Scheme~1--4 at the final decision epoch ($t=15$) for  system VoS and over the evaluation horizon for requested-target recall and RMSE. The Oracle values serve as reference scores.}
	\label{tab:closed_loop_schemes}
	\footnotesize
	\renewcommand{\arraystretch}{1.15}
	\begin{tabular*}{\textwidth}{@{\extracolsep{\fill}}lccccc@{}}
		\hline
		Metric & Scheme~1 & Scheme~2 & Scheme~3 & Scheme~4 & Oracle \\
		\hline
		System VoS ($t=15$)
		& $0.634 \pm 0.004$ & $0.600 \pm 0.005$ & $0.568 \pm 0.005$
		& $0.514 \pm 0.006$ & $0.660$ \\
		
		Horizon-average target recall
		& $0.896 \pm 0.009$ & $0.810 \pm 0.011$ & $0.752 \pm 0.012$
		& $0.622 \pm 0.014$ & $1.000$ \\
		
		Horizon RMSE (m)
		& $1.14 \pm 0.04$ & $1.33 \pm 0.04$ & $1.47 \pm 0.04$
		& $1.78 \pm 0.05$ & $0.90$ \\
		\hline
	\end{tabular*}
\end{table*}

\subsection{Evaluation Setup and Comparison Schemes}

We run the simulations in MATLAB 2025b on a laptop with an Apple M1 chip and 8 GB of memory. The transmitter and receiver have $N_t=8$ and $N_r=8$ antennas, respectively. The total bandwidth is $B_{\max}=100$ MHz, with $B_0=180$ kHz per subchannel. The default maximum transmit power is $P_{\max}=10$ dBW, and the maximum number of iterations for the iterative optimization procedure is $J_{\max}=150$. The communication and sensing service weights are randomly generated within $[0,1]$, while the sensing requirements and communication data-rate requirements are drawn from $[0.01,10]$ and $[0,100]$, respectively. The objective-function parameters are also randomly generated to capture heterogeneous service demands. Unless otherwise specified, we set $\alpha=\beta=0.8$. We average each metric over 100 independent runs with random service requests and network configurations. We compare four resource allocation schemes.

\textit{Scheme~1 (Proposed)} uses the supporting-price allocator in
Algorithm~\ref{alg:sg_online} to select subchannel assignments and
optimize transmit power. It warm-starts the search from the screened
assignment of the previous epoch when feasible. A single sensing
operation for a common target contributes to the VoS of every agent
requesting that target. \textit{Scheme~2}~
\cite{11653527} assigns operations to idle
subchannels one at a time. At each step, it keeps earlier assignments
fixed and reoptimizes transmit power for each candidate assignment. \textit{Scheme~3} uses the same search as Scheme~1 but assigns
a separate sensing operation to each agent--target request. Requests
from different agents for the same physical target are thus served
through separate allocations. \textit{Scheme~4} assigns subchannels cyclically to communication and sensing operations and allocates equal transmit power across active
subchannels.


\subsection{VoS under Different Network Conditions}

Fig.~\ref{fig:sim_resources}(a) shows that the system VoS increases with the transmit power budget, as the additional power provides greater flexibility to accommodate heterogeneous communication and sensing requirements. The proposed scheme consistently achieves the highest VoS among the considered schemes. The supporting-price mechanism coordinates the competing communication and sensing demands.

In Fig.~\ref{fig:sim_resources}(b), as the common-request probability increases, the system VoS improves across all schemes because overlapping sensing requests create more opportunities to share sensing results and avoid redundant resource consumption. The proposed scheme obtains a more pronounced benefit from these opportunities, maintaining a performance advantage over the benchmarks. The benefit of shared sensing thus depends on how often agents request the same targets.

In Fig.~\ref{fig:sim_resources}(c), the system VoS decreases as more agents with heterogeneous demands compete for limited resources. Nevertheless, the proposed scheme consistently achieves higher VoS than the benchmarks. This advantage suggests that the proposed allocation mechanism is more effective in coordinating competing demands and preserving service fulfillment as the network becomes increasingly crowded.

\begin{figure*}[htbp]
	\centering
	\includegraphics[width = 0.95 \textwidth]{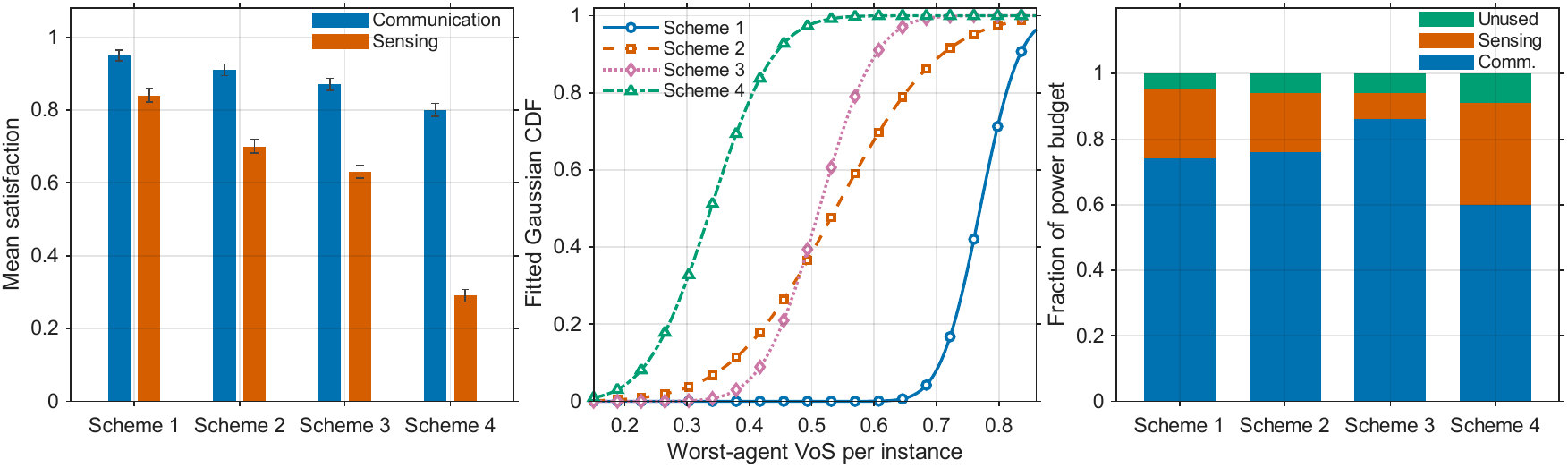}
	\caption{Service satisfaction and power allocation. (a) Mean communication and sensing satisfaction with comparison ranges. (b) Gaussian CDF curves of the minimum individual VoS per instance. (c) Proportions of the power budget allocated to communication and sensing, alongside the unused share.}
	\label{fig:sim_service}
\end{figure*}

\begin{figure*}[htbp]
	\centering
	\includegraphics[width= 0.95 \textwidth]{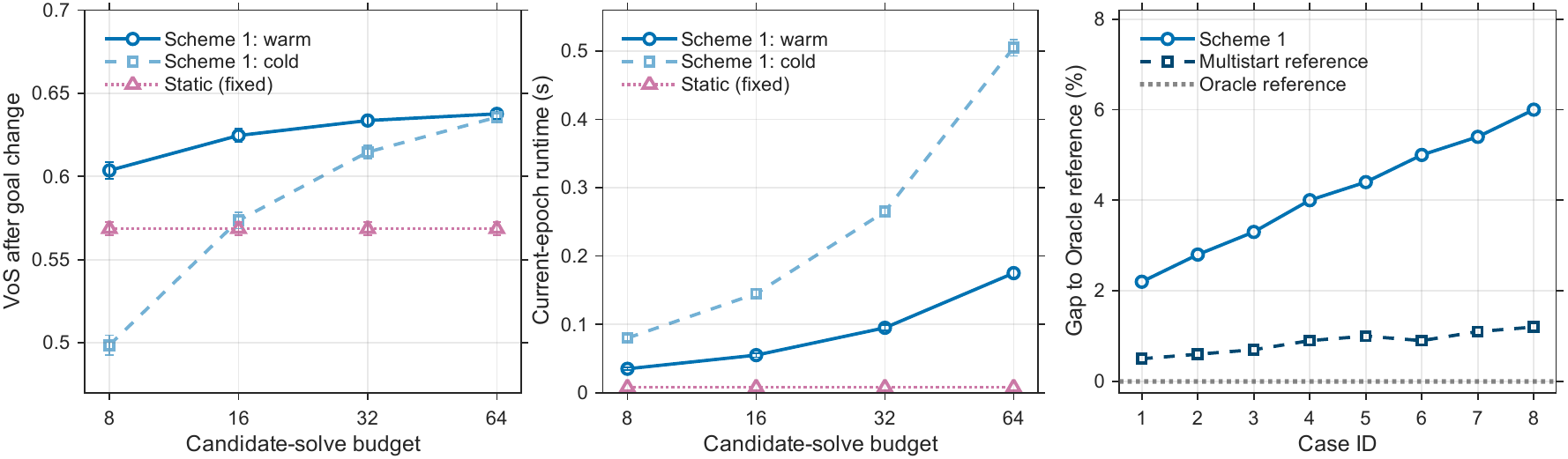}
	\caption{Performance comparison following a goal change: (a) Post-change VoS and (b) runtime overhead versus the candidate-solve budget. (c) VoS gaps relative to the case-specific Oracle references. Error bars indicate comparison ranges.}
	\label{fig:sim_adaptation}
\end{figure*}

\subsection{Service Satisfaction and Resource Utilization}

Fig.~\ref{fig:sim_service}(a) compares the average satisfaction levels in the two service dimensions. The proposed scheme achieves high satisfaction in both, indicating that its VoS-oriented allocation can accommodate communication and sensing requirements jointly. By comparison, the benchmarks exhibit more pronounced degradation, particularly in sensing satisfaction. This difference suggests that optimizing resource allocation without explicitly accounting for the joint service objective may leave one service dimension insufficiently supported, even when the other achieves relatively high satisfaction.

Fig.~\ref{fig:sim_service}(b) shows the CDF of the minimum individual VoS across instances. The proposed scheme produces a distribution shifted toward higher values, indicating that more agents achieve better minimum service fulfillment. The improved lower tail indicates that instances with very low minimum individual VoS are less frequent.

Fig.~\ref{fig:sim_service}(c) shows the proportions of the power budget allocated to communication, sensing, and left unused. The proposed scheme allocates resources to both service dimensions while keeping the unused share relatively small. Importantly, the results indicate that allocating a larger resource share to a particular service does not necessarily yield higher satisfaction in that dimension. The resulting satisfaction instead depends on how well the allocation matches the corresponding service requirements. This explains why the proposed scheme achieves higher overall VoS through demand-aware resource coordination, rather than simply favoring communication or sensing.

\subsection{Adaptation to Changing Service Goals}

Fig.~\ref{fig:sim_adaptation}(a) reports the post-change system VoS under different candidate-solve budgets. The proposed algorithm achieves higher VoS as the budget increases, since more candidate allocations can be evaluated to refine the updated solution. Moreover, warm-start consistently provides better performance than cold-start under limited budgets. By reusing the previous allocation as an informative initialization, warm-start can exploit the structure of the preceding solution and focus the available computational effort on adapting to the changed goals. As the budget increases, the performance gap narrows, suggesting that cold-start can partially compensate for its less informative initialization.

Fig.~\ref{fig:sim_adaptation}(b) compares the corresponding runtime. For both strategies, the runtime increases with the candidate-solve budget because more candidate solutions require evaluation. However, warm-start consistently incurs lower computational overhead than cold-start. Together with Fig.~\ref{fig:sim_adaptation}(a), this shows that warm starts improve post-change VoS under limited budgets while reducing runtime.

Fig.~\ref{fig:sim_adaptation}(c) reports the relative VoS gaps to the case-specific Oracle and multistart references. The proposed scheme maintains relatively small gaps to these references, and a larger candidate-solve budget generally reduces the gaps. This trend confirms that additional candidate evaluations can improve the obtained allocation, while the warm-start strategy offers an efficient means of approaching the reference performance. These comparisons characterize solution quality in the tested instances and do not establish global optimality for the numerically obtained solutions.
\subsection{Closed-Loop Feedback Performance}
\begin{figure*}[htbp]
	\centering
	\includegraphics[width= \textwidth]{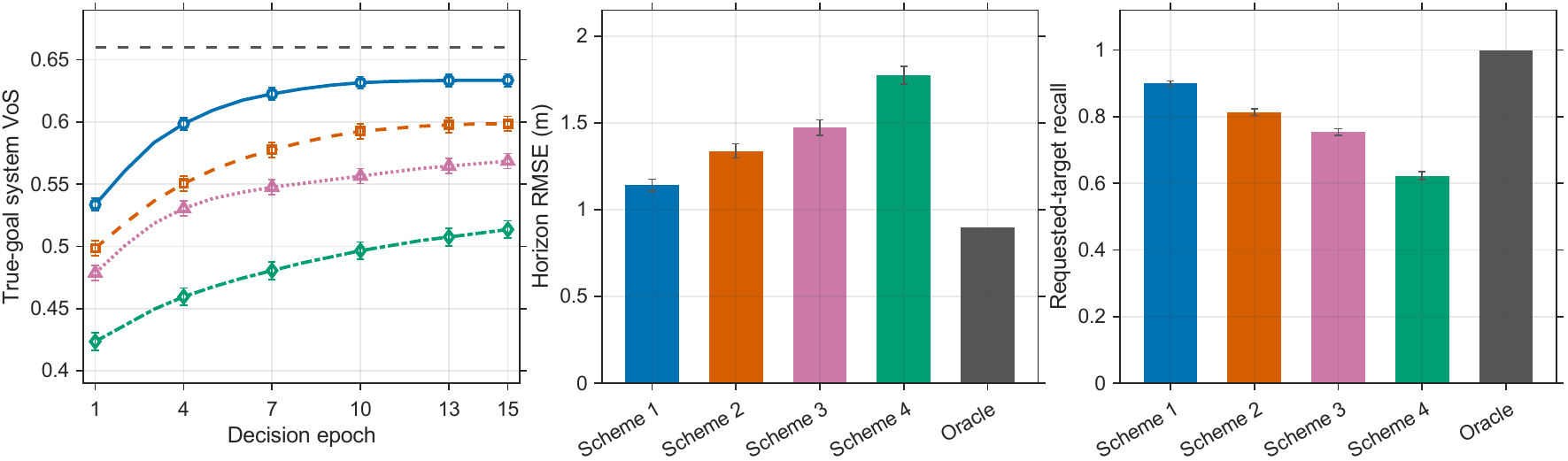}
	\caption{Closed-loop feedback performance over 15 decision epochs: (a)  VoS trajectories, (b) horizon RMSE, and (c) horizon-average requested-target recall. Error bars indicate comparison ranges. The corresponding aggregate results are  in Table~\ref{tab:closed_loop_schemes}.}
	\label{fig:sim_feedback}
\end{figure*}
Fig.~\ref{fig:sim_feedback}(a) shows the system VoS against the true service goals over 15 epochs of closed-loop operation. The proposed scheme maintains higher VoS than the benchmarks throughout the considered horizon, with its performance improving as feedback accumulates. This trend indicates that the framework can progressively exploit updated goal information to refine resource allocation, rather than relying on a fixed estimate of agents' demands. At the final epoch, Scheme~1 achieves a system VoS of 0.634, compared with 0.600, 0.568, and 0.514 for Schemes~2--4, respectively, as reported in Table~\ref{tab:closed_loop_schemes}. The Oracle, with access to the true service goals, achieves 0.660.

Fig.~\ref{fig:sim_feedback}(b) reports the RMSE of the estimated service-goal horizon. The proposed scheme achieves lower RMSE than the benchmarks, indicating more accurate estimation of future service goals. As summarized in Table~\ref{tab:closed_loop_schemes}, its horizon RMSE is 1.14 m, compared with 1.33 m, 1.47 m, and 1.78 m for Schemes~2--4, respectively. The lower estimation error suggests that the feedback mechanism provides more informative estimates for subsequent allocation decisions, helping the algorithm anticipate changes in agents' requests.

Fig.~\ref{fig:sim_feedback}(c) presents the horizon-average requested-target recall. Scheme~1 achieves a recall of 0.896, outperforming Schemes~2--4, which obtain 0.810, 0.752, and 0.622, respectively. This improvement indicates more effective identification and tracking of agents' requested sensing targets. Since missed targets can limit sensing service fulfillment, the higher recall helps explain the higher VoS in Fig.~\ref{fig:sim_feedback}(a).

\section{Conclusion}
\label{sec:conclusion}

We developed a framework for intent-driven ISAC in which
communication and sensing outcomes inform subsequent resource
decisions. Agents translate persistent user intents into service
goals that evolve with task progress and new observations. The
VoS measures how well the current goals are met
and provides the objective for joint subchannel and power allocation.
A shared localization result is evaluated against each requesting
agent's accuracy requirement and target relevance, allowing the
allocation to account for the benefit of sensing across tasks. To coordinate contributions to shared sensing, we derived
supporting prices that deduct the marginal VoS delivered to other
agents from the common power price. With admissible supporting
prices and optimistic equilibrium selection, the Stackelberg
formulation and the centralized per-epoch allocation problem have
the same optimal VoS. The online allocator searches neighboring
subchannel assignments, optimizes the associated powers, and
screens candidates using bounds on unilateral deviation gains. 


\bibliographystyle{IEEEtran}
\vspace{-3mm}
\bibliography{references}

\end{document}